\documentclass[12pt,preprint]{aastex}

\begin{document}

\title{Dynamical Masses of Young Stars in Multiple Systems}

\author{G. H. Schaefer, M. Simon, }
\affil{Department of Physics \& Astronomy, SUNY Stony Brook, NY 11794-3800}

\author{E. Nelan, and S. T. Holfeltz}
\affil{Space Telescope Science Institute, Baltimore, MD 21218}

\begin{abstract}

We present recent measurements of the orbital motion in the young binaries DF Tau and ZZ Tau, and the hierarchical triple Elias 12, that were obtained with the Fine Guidance Sensors on the HST and at the Keck Observatory using adaptive optics.  Combining these observations with previous measurements from the literature, we compute preliminary orbital parameters for DF Tau and ZZ Tau.  We find that the orbital elements cannot yet be determined precisely because the orbital coverage spans only $\sim$90$\degr$ in position angle.  Nonetheless, the range of possible values for the period and semi-major axis already defines a useful estimate for the total mass in DF Tau and ZZ Tau, with values of (0.90$^{+0.85}_{-0.35}$) M$_\sun$ and (0.81$^{+0.44}_{-0.25}$) M$_\sun$, respectively, at a fiducial distance of 140 pc.

\end{abstract}

\section{Introduction}

The masses and ages of pre-main sequence stars are commonly determined from their position relative to theoretical tracks of stellar evolution on a color-magnitude diagram. This method of mass determination is subject not only to uncertainties in the measured luminosity and effective temperature of the star but also to discrepancies among evolutionary models. These discrepancies arise from differences in atmospheric opacities and theories of interior convection.  Currently there are few reliably measured masses for low-mass (M $<$ 1 M$_\sun$) pre-main sequence stars (Simon, Dutrey, \& Guilloteau 2000).  The determination of stellar masses through dynamical techniques provides a test for evolutionary models of young stars.

Observations of a resolved binary reveal the projected separation of the system and the position angle of the secondary with respect to the primary.  With sufficient orbital coverage, the orbital elements of the system can be computed (e.g. Couteau 1981).  If the distance to the binary system is known, the total mass can also be determined.  The relative orbit of a visual binary alone cannot yield the masses of the individual components.  To achieve this, one also has to determine the mass ratio, either by observing the motion of the components around their center of mass or by measuring their radial velocities relative to the center of mass velocity.

In this paper we present new positional measurements for two classical T Tauri binary systems DF Tau and ZZ Tau and the hierarchical triple Elias 12, all located in the Taurus star-forming region.  Most of the measurements we report here were taken with the Fine Guidance Sensors on the Hubble Space Telescope (HST) and are a continuation of the program described in Simon, Holfeltz, \& Taff (1996).  The close pair in Elias 12 was also observed with adaptive optics on the Keck II telescope.  Thi\'{e}baut et al. (1995) and Tamazian et al. (2002) estimate the orbital parameters for DF Tau.  We performed preliminary orbital calculations for DF Tau and ZZ Tau, and find that we have not yet observed enough of an orbit for these systems to determine the orbital parameters precisely.  However, although the range of probable values for the parameters is large, the relation between the period and semi-major axis, through Kepler's Third Law, limits the total mass of these systems with sufficient precision as to be useful.

\section{Observations}

\subsection{Fine Guidance Sensor Observations}

The Fine Guidance Sensors (FGS) are a set of three optical interferometers on board the HST (Nelan \& Makidon 2001).  Each FGS consists of a pair of Koesters prisms that detect the tilt of the incoming wavefront in two orthogonal directions.  The photon counts are measured by pairs of photomultipliers with an S-20 response.  The normalized difference between the intensities of the two beams emerging from each prism, as a function of the tilt of the incoming wavefront, produces the interferometric fringe pattern of the target, also known as the FGS transfer function.  

The interferometric response of a binary system is the linear superposition of two single star transfer functions, scaled by the relative brightness of the components and shifted according to their separation on the sky.  The transfer function of a binary can be deconvolved by modeling the system with the observation an unresolved single star.  The shape of the fringe pattern depends on the color of the target.  Therefore, we use the FGS single star calibrators SAO 185689 and HD 233877, whose colors closely match those of DF Tau, ZZ Tau, and Elias 12 (listed in Table 1).  To model the FGS observations of binaries we followed a least-squares approach of searching through a grid of separations and brightness ratios for the components (Lattanzi et al. 1992) and finding the minimum in the $\chi^2$ surface.  The software we developed to analyze triples follows the same method, but over grids of separation and brightness ratios for three stars.  The separations and their uncertainties derived from our program, when applied to binary systems, are consistent with those obtained using the software at the Space Telescope Science Institute. 

The refurbished FGS1r was installed during the HST servicing mission of 1997 and later became the preferred science instrument because of its improved angular resolution (Nelan \& Makidon 2001).  The FGS observations taken before June 1999 were measured with FGS3 (e.g. Holfeltz et al. 1995) with the PUPIL filter.  Following this date, the targets were observed with FGS1r with the F583W filter and the FGS in its high angular resolution transfer mode.

Each observation of the target was obtained during one HST orbit and was composed of multiple ($>$10) interferometric scans.  Scan lengths of 1.0$''$ to 1.5$''$ were used to measure the FGS baseline on either side of the binary.  The x and y scans were each cross-correlated, co-added, and smoothed by piecewise polynomial fits to retain the interferometric signal but remove the statistical noise of counting photons.  Figure 1 shows these steps in processing applied to ZZ Tau as observed along the x-axis of the FGS on 6 December 2002.

Table 1 lists our most recent FGS measurements for DF Tau, ZZ Tau, and Elias 12.  This table also includes previous results from the literature for DF Tau (Chen 1990; Thi\'{e}baut et al. 1995; Ghez et al. 1995; Simon et al. 1996; White \& Ghez 2001; and Balega et al. 2002), ZZ Tau (Simon et al. 1996), and Elias 12 (Ghez et al. 1995; Simon et al. 1996; and White \& Ghez 2001).  Column (2) specifies the photometric band for the primary and secondary magnitudes listed in columns (3) and (4).  The separation and position angle are given in columns (5) and (6) respectively.  The FGS observations of Elias 12 were modeled as a triple system (S-Na-Nb).  In Table 1, we present our measurements of the relative position of the wide pairs Elias 12 S-Na and S-Nb, as well as the relative position measured for the close pair Elias 12 Na-Nb.  

The speckle observations of Thi\'{e}baut et al. (1995) and Balega et al. (2002), listed in Table 1, indicate an orientation for DF Tau that is rotated by 180$\degr$.  This discrepancy in the identification of the primary may be attributable to the spectral energy distributions of the components as well as to the variability of the system.  For the purpose of computing the orbital parameters, we add 180$\degr$ to these position angle measurements.  For the purpose of computing the orbital parameters, we add 180$\degr$ to these position angle measurements.  All of the observations of ZZ Tau prior to 1997.71 had projected separations along the x-axis of the FGS that were smaller than 20 milliarcseconds (mas).  Measuring separations this small with FGS3 was not reliable, therefore we determined the position of the binary system from the y-axis measurements of consecutive observations. Assuming that the stars have not moved significantly between pairs of observations and since the orientation of the y-axis was sufficiently different from one observation to the next, we solved for the (x,y) position at the intermediate date, halfway between the observations.  The reported observation on 1995.08 is a combination of the measured y-axis separations on 1994.93 and 1995.23.  Likewise, the 1996.12 results were obtained by merging the observations on 1995.99, 1996.15, and 1996.24.  The average orbital motion of $\sim$ 9 mas/year for subsequent reliable observations justifies the assumption that the stars have not moved significantly between the dates used to calculate the effective points.  Moreover, these effective points are consistent with the later FGS1r data.

The V-band magnitude of a target observed with the FGS is given by
\begin{equation}
V = -2.5 \log{(C)} + \alpha (B-V) + \beta ~ mag.
\label{Vmag}
\end{equation}
where C is the sum of the counts (per 25 msec) detected by the photomultipliers along the orthogonal x and y axes and $(B-V)$ is the color index of the target.  For the FGS3 with the PUPIL filter, the calibration coefficients are $\alpha = 0.32 \pm 0.31$ and $\beta = 19.2 \pm 0.1$ (Holfeltz et al. 1995).  The magnitudes for DF Tau and Elias 12 reported by Simon et al. (1996) and listed in Table 1 have been adjusted according to recent measurements of the $(B-V)$ colors (White \& Ghez 2001, see Table 1).  For Elias 12, we use a value of $(B-V) \sim 1.2$ for the system, which is the average color of the northern and southern components.

The imprecision of the $\alpha$ and $\beta$ coefficients for FGS3 contributes to a systematic uncertainty of $\sim$0.5 mag in the V-band for a typical value of $(B-V)$ = 1.5 mag.  The photometric calibration for FGS1r is also subject to similar systematic uncertainties that produce a constant offset in the magnitude at a specified $(B-V)$.  To avoid an artificial jump in the magnitude measured by the old and new FGS, we applied an offset to the FGS1r results for each system, so that the average magnitudes of the non-varying components are in agreement with the earlier results obtained with FGS3.  The FGS magnitude errors reported in Table 1 correspond to C$^{1/2}$ counting statistics and do not include the systematic uncertainty in the $\alpha$ and $\beta$ coefficients.  The magnitude errors for the close pair in Elias 12 also includes the uncertainties in the flux ratio derived from the model fits.

\subsection{Near-IR Adaptive Optics Observations}

Elias 12N was resolved as a close binary through a lunar occultation observation (Simon et al. 1995).  This close pair is located $\sim$300 mas away from the primary Elias 12S.  Since the wide member of the triple system is bright and lies within the isoplanatic patch of the close pair, the primary provides a suitable point-spread function (PSF) for measuring the separations and flux ratios of the close pair through high resolution imaging.

The hierarchical triple Elias 12 was observed with the near-infrared camera NIRC-2 at the Keck II telescope using the adaptive optics (AO) system (Wizinowich et al. 2000) on two occasions.  The NIRC-2 detector is a 1024$\times$1024 Alladin-3 InSb array.  The images were taken with the H and K$'$ filters, with a pixel scale of 0.01$''$/pixel.  Each image is composed of 10 co-adds.  Sets of five images were obtained using a five-point dither pattern with a leg size of 2$''$.  The observations on 8 March 2002 have an exposure time of 0.25 seconds and an AO rate of 272 Hz.  On 30 October 2002 we observed with an exposure time ranging from 0.2-0.5 seconds and an AO rate of 120-140 Hz.  During both times of observation, Elias 12S was used as the guide star.

The NIRC-2 images were re-binned to expand the pixel scale from 0.010$''/$pixel to 0.002$''$/pixel using bilinear interpolation to improve the positional accuracy.  Subarrays with a width of 200 pixels were extracted from the expanded image; one array was centered on the primary and the other was centered on the close pair.  The PSF was used to construct models of the close pair through a grid of separations and flux ratios.  The positions of the companions in the close pair were determined from where the variance between the model and the observation reached a minimum.  Figure 2 shows the model for the H-band observations obtained on 30 October 2002.  The positional results are presented in Table 1.  The flux ratios of the components in Elias 12 in the H- and K$'$-bands are presented in Table 2.  For comparison, the flux ratios in the V-band, measured by the FGS, are also included.  Column (3) lists the ratio of each component in the close pair in relation to the widely separated primary, Elias 12S.  Column (4) lists the flux ratio of the close pair, (Nb/Na). 

\section{Computation of Orbital Elements}

We calculated the orbital elements following the three dimensional grid search procedure outlined by Hartkopf, McAlister, \& Franz (1989) and Mason, Douglass, \& Hartkopf (1999).  The program searches through a grid of solutions formed by varying the initial estimates for the period, time of periastron passage, and eccentricity.  The program locates the solution where the $\chi^2$ between the measured and calculated positions at the times of observation is at a minimum.
 
The initial values for the orbital elements were obtained through a geometric approach based upon a preliminary fit to the apparent ellipse.  Interpolating curves were drawn for the position angle PA(t) and the separation $\rho$(t) (Aitken 1964).  These curves were constructed to vary smoothly with time and to yield a consistent value for the constant of areas, $\rho_{ave}^2 \frac{\Delta (PA)}{\Delta t}$, across equally spaced time intervals.  For each date of observation, values for the position angle and separation were read from the interpolating curves to provide a set of position measurements adjusted to implicitly include the time component of the orbital motion.  The apparent orbit was determined from the interpolated data points by performing a least-squares fit to an ellipse (Haralick 1992).   The period, time of periastron passage, and eccentricity (P,T,e) were calculated from the geometry of the apparent ellipse (Couteau 1981).  The interpolated position measurements were used only to determine the initial estimates.

The initial estimates for the orbital elements were adjusted through the three dimensional grid search procedure.  Input to the program includes a set of starting values (P,T,e) and the corresponding step sizes and search ranges for each of the three parameters.  For each iteration through the grid, a linear least-squares fit is performed to determine the Thiele-Innes elements (A,B,F,G).  The angular semi-major axis, inclination, position angle of the line of nodes, and angle between the node and periastron (a,i,$\Omega$,$\omega$) are computed from the Thiele-Innes elements.  The data are weighted by their respective measurement errors and the $\chi^2$ between the calculated and observed positions is computed.  The set of parameters that produces a minimum in the $\chi^2$ surface is retained as the orbital solution.

Formal errors in the orbital elements are derived using the general matrix method for multiple regression (Bevington \& Robinson 1992).  These error estimates involve linearizing the equations of orbital motion through a Taylor expansion terminated after the first order.  For orbits that are poorly determined, higher order terms may become important.  For such cases, it is necessary to examine the shape of the $\chi^2$ surface.  The precision of the computed orbital elements can be evaluated from the distribution of orbital elements that produces a variation of one within the reduced $\chi^2$.

Hartkopf, Mason, \& Worley (2001) describe a scheme for evaluating the reliability of the observed orbit.  The orbits are graded on the basis of orbital coverage, the quality of the measurements, and the number of observations.  The grades are assigned on a numerical scale, with 1 being definitive and 5 considered indeterminate.  We tested our program on published orbits from each of the five numerical categories (Mason et al. 2001).  The errors were computed through the formal technique as well as through the $\chi^2$ analysis.  For the orbits chosen with grades 1-3, the orbital elements computed from our program agree with the published values, within $1\sigma$.  The size of both the formal and $\chi^2$ analysis errors are consistent with the published errors.  For the orbits of grades 4 and 5, the computed and published preliminary orbital elements were discrepant by as much as 3$\sigma$, when considering the formal errors.  If the errors from the $\chi^2$ analysis are used, the computed results agree with the published values within $1\sigma$.  Therefore, when determining the precision of the computed orbital elements for low grade orbits, it is important to consider the shape of the $\chi^2$ surface.

\section{Orbital Parameters}

\subsection{DF Tau}

The orbital motion of DF Tau, using the data in Table 1, is shown in Figure 3.  The initial estimates for the orbital parameters of DF Tau (P=44 years, T=1982, e=0.23) were obtained through the geometric analysis of the preliminary apparent ellipse.  Although these values seemed reliable, we found, upon analysis using the grid search procedure, that the $\chi^2$ surface is broad and flat in the region of the mimimum.  Consequently, the initial estimates were not critical to the final analysis.  The range of possible solutions extends from 30 to 3000 years in period and from 0.08$''$ to 1.1$''$ in semi-major axis, all within a variation of 1 within the minimum reduced $\chi^2$.  The inclination is well-defined across all possible periods, with a value of 132$\degr \pm$13$\degr$.  Figure 4 shows three examples of possible orbital solutions for DF Tau at periods of 50, 100, and 150 years, which agree within $\Delta (\chi^2/ \nu) < 0.01$, where $\nu$ is the number of degrees of freedom.  The broad range of parameters calculated through the $\chi^2$ analysis includes the values of the orbital elements determined by Thi\'{e}baut et al. (1995) and Tamazian et al. (2002).  


Our orbital calculations for DF Tau indicate that the system has not been observed with sufficient coverage to determine the orbital elements definitively.  This is consistent with its designation as a grade 4 orbit in The Sixth Catalog of Orbits of Visual Binary Stars (Hartkopf \& Mason 2002).  This grade assignment indicates that any orbital determination should be considered preliminary, where substantial revision may be required for the individual elements.  This agrees with our results.

\subsection{ZZ Tau}

Figure 5 shows the orbital motion of ZZ Tau, as measured with the FGS.  The grid search procedure indicates a large variation in the orbital elements computed for ZZ Tau.  The region within $\Delta (\chi^2/ \nu) = 1$ on the $\chi^2$ surface reveals that orbits with periods in the range of 20 to 1500 years and semi-major axes from 0.05$''$ to 0.66$''$ fit the measured data with similar accuracy.  Through a broad search across all possible periods, the inclination is again well-defined, i = 118$\degr \pm$7$\degr$.  Three examples of possible orbital solutions for ZZ Tau are plotted in Figure 5.  These solutions have periods of 25, 50, and 75 years and all agree within $\Delta (\chi^2/ \nu) < 0.05$.  


\subsection{Elias 12}

We measured the orbital motion of the three components in the hierarchical triple Elias 12.   Figure 6 shows the positions of Elias 12 Na and Nb relative to Elias 12S.  Figure 7 shows the orbital motion observed for the close pair, where the position of Elias 12 Nb is plotted relative to Elias 12Na.  We have not observed the orbital motion in the close pair, Elias 12 Na-Nb, with sufficient coverage to justify calculating orbital solutions at the current time.  However, over the past three years Elias 12Nb has moved $\sim$50$\degr$ with respect to Elias 12Na, so a preliminary orbital solution might be possible in the near future.  Additionally, Elias 12S provides a phase reference to measure the motion of the close pair about their center of mass.  This will provide an estimate of the mass ratio, and hence the individual masses of Elias 12 Na and Nb.  

\section{Masses}

The total mass of a binary is determined by a$^3$/P$^2$, with the semi-major axis on a physical scale.  We use the measurement of the average distance to the Taurus cluster, d=140 pc, from Kenyon, Dobrzycka, \& Hartmann (1994) to convert the angular separation to a physical length.  Figure 8 shows the distribution of possible values for the total mass, M$_{tot}$ = (a$^3$/P$^2$)(d/140pc)$^3$ M$_\sun$, for DF Tau.  The individual values were obtained by searching for orbital solutions, within $\Delta (\chi^2/ \nu) = 1$ from the minimum $\chi^2$, across the broad range of permitted periods for each system, eccentricities from 0.0 to 0.9, and times of periastron passage between 1900-2100.  We also required that the total mass of the system be less than 4 M$_\sun$ since masses this large would clearly not be representative of the spectral types and luminosities of the stars observed within the Taurus star-forming region.  This condition eliminated highly eccentric, short period systems from the sample of possible solutions.  

Although the observations allow a broad range of orbital parameters, Figure 8 shows that the total mass of DF Tau is well-defined at M$_{tot}$ = (0.90$^{+0.85}_{-0.35}$)(d/140pc)$^3$ M$_\sun$.  Since the mass distribution is asymetrical, the reported value for the system mass was calculated using the median of the distribution.  The range of the quoted errors include 34\% of the values on each side of the median.  A similar analysis for ZZ Tau yields M$_{tot}$ = (0.81$^{+0.44}_{-0.25}$)(d/140pc)$^3$ M$_\sun$.  The result that poorly determined orbits can still produce useful values for the total mass is also noted by Eggen (1967) and Hartkopf \& Mason (2002).  The system mass is determined by the instantaneous radial acceleration and separation of the stars.  Although the complete orbit may not be well-determined, the orbital solutions must reproduce the projected radial acceleration of the system at the current separation, measured through the curvature of the apparent orbit.  As a result, the value of (a$^3$/P$^2$) can be determined with much greater accuracy than the uncertainty in the orbital elements suggests.

From angularly resolved HST/STIS spectra of DF Tau, Hartigan \& Kenyon (2003) found that the spectral types of the primary and secondary are M2.0 and M2.5, respectively.  Our value of M$_{tot}$, and the masses implied by these spectral types, favor the evolutionary tracks calculated by Baraffe et al. (1998) or Pahler \& Stahler (1999) over those of D'Antona \& Mazzitelli (1997).  This is consistent with the earlier finding for the masses of single stars in the mass range 0.7 to 1 M$_\sun$, measured from the rotation of circumstellar disks (Simon et al. 2000), and extends the range down to $\sim$0.4 M$_\sun$.  The spectral type of ZZ Tau, observed as an unresolved system, is M3 (Herbig \& Bell 1988).  Our estimate of M$_{tot}$ for ZZ Tau is also consistent with the Baraffe et al. (1998) tracks if the components are of equal mass.

The precision of the total mass measurements is expected to improve rapidly with continued observations over the next few years.  To explore this, we assumed orbital solutions with a period of 50 years that satisfy the data presently available for each system, DF Tau and ZZ Tau.  We then calculated their relative positions in the future at annual intervals, applied the $\pm$2 mas uncertainty characteristic of FGS observations, and solved for the orbital parameters.  For a period of 50 years, we found that it would take $\sim$15 years of observations into the future, from the year 2003, to determine reliable orbital parameters for DF Tau and $\sim$20 years for ZZ Tau.  We also found, however, that for each system, annual observations for another 5 years will decrease the uncertainty in the M$_{tot}$ estimate by a factor of $\sim$2, a significant advance and useful for discriminating among the theoretical evolutionary tracks.  The expected improvement in the total mass estimate, over the next 5 years, for DF Tau is shown in Figure 8.  The simulated mass distributions in this figure were obtained for assumed periods of 50 and 100 years.  

A meaningful test of the available pre-main sequence evolutionary tracks requires individual masses measured with $\sim$5\% absolute precision.  Masses determined by mapping orbits scale with the distance as d$^3$.  The Taurus star-forming region subtends $\sim$10$\degr$ on the sky.  If its depth is comparable to its width, then the distance uncertainty to any given member is $\pm$10 pc, which produces a 20\% uncertainty in the system mass.  The required precision in the mass could be reached by a direct parallax measurement with $\pm$0.05 mas accuracy.  This would determine the distance to a star in Taurus to $\pm$1 pc, providing a 2\% system mass measurement.  Measuring a parallax with this accuracy and precision is well within the capability of the Space Interferometry Mission.

\section{Photometric Variability of DF Tau}

The brightness of DF Tau varies with a period of $\sim$8.5 days (Bouvier \& Bertout 1989).  Bouvier et al. (1993) interpret these fluctuations as the rotational modulation of the flux by a hot spot on the stellar surface.  Li et al. (2001) observed the spectroscopic signature of mass accretion from line profiles over nearly a complete cycle of a flare event of DF Tau.  

With the FGS we are able to obtain the magnitudes for each of the resolved components in DF Tau.  Our measurements indicate that the primary varies with an amplitude of $\sim$1.5 mag in V, as shown in Figure 9.  The bottom panel of Figure 9 shows that the brightness of the secondary remains constant in comparison to the fluctuations of the primary.  The amplitude of variation observed in the primary is consistent with previously published results obtained for the unresolved system (Lamzin et al. 2001; Li et al. 2001; Herbst et al. 1994; Bouvier et al. 1993; and references therein).  The photometric variations, as measured by the FGS, indicate that the observed rotational modulation applies to the primary, although the FGS data were not obtained frequently enough to measure the periodicity.

Lamzin et al. (2001) analyze the historical lightcurve of DF Tau over the past century.  They suggest that the long-term photometric variations of $\sim$40 years may be caused by circumstellar disk accretion on to the primary, modulated by the orbital motion of the companion.  This result was obtained by considering the orbital period of 82$\pm$12 years reported by Thi\'{e}baut et al. (1995).  Although a period of 80 years is included in the uncertainty range of our orbital analysis, we find that the period is not known with sufficient reliability to confirm the effect of orbital modulation.

\section{Summary}

The main results of our preliminary orbital calculations, based upon measurements from the literature and our observations with the FGS of the HST and adaptive optics at the Keck Observatory, are as follows.

1. The orbits of DF Tau and ZZ Tau have not been observed with sufficient orbital coverage to determine the orbital parameters precisely.

2. Although the observations allow a broad range of orbital parameters, we are able to obtain useful values for the total dynamical masses of DF Tau and ZZ Tau, with values of (0.90$^{+0.85}_{-0.35}$) M$_\sun$ and (0.81$^{+0.44}_{-0.25}$) M$_\sun$, respectively, at an estimated distance of 140 pc.  The precision in these mass estimates is expected to improve by a factor of $\sim$2 with continued observations over the next five years.

3. Analysis of the relative orbital motion observed in the close pair of the hierarchical triple Elias 12 will become possible as more of the orbit is measured in the coming years.  The northern components are currently separated by $\sim$40 mas.

\acknowledgements

We thank W.I. Hartkopf for providing his binary orbit program, L. Prato for helping with the adaptive optics observations at the Keck Observatory, and D. Peterson for advice on computing errors in the orbital parameters.  We also thank W. Van Altena for leading us to the Eggen reference and the anonymous referee for providing helpful comments.  This research is based on observations made with the NASA/ESA Hubble Space Telescope, obtained at the Space Telescope Science Institute, which is operated by the Association of Universities for Research in Astronomy, Inc., under NASA contract NAS 5-26555.  These observations are associated with GO Proposals 6486, 7487, 8339, 8616, 8783, 9229, and 9335.  Some of the data presented herein were also obtained at the W.M. Keck Observatory, which is operated as a scientific partnership among the California Institute of Technology, the University of California, and the National Aeronautics and Space Administration. The Observatory was made possible by the generous financial support of the W.M. Keck Foundation.  We wish to recognize and thank the Hawaiian community for the opportunity to conduct these observations from the summit of Mauna Kea.  This research has made use of the Washington Double Star Catalog maintained at the U.S. Naval Observatory.  


\clearpage

\begin{deluxetable}{lclllll} 
\tablewidth{0pt}
\tablecaption{Relative Positions of Pre-Main Sequence Stars in Multiple Systems} 
\tablehead{
\colhead{Year} & \colhead{Band} & \colhead{m$_A$}   & \colhead{m$_B$}   &
\colhead{$\rho$(mas)} & \colhead{P.A.($\degr$)} & \colhead{Ref.} }
\startdata 
\cutinhead{DF Tau (HBC 36), (B-V) = 1.6\tablenotemark{a} mag}
1986.800 & K & 7.47$\pm$0.02 & 7.88$\pm$0.02 & 73.0$\pm$13 & 350.0$\pm$3.0  & 1 \\
1989.840 & 656 nm & &  & 82$\pm$2 & 167.0$\pm$3.0 & 2 \\
1990.857 & K & & & 90.0$\pm$2  & 328.0$\pm$3.0 & 3 \\ 
1991.730 & 700 nm & & & 84$\pm$4 & 142.0$\pm$1.0 & 2 \\
1992.778 & H & & & 98.0$\pm$8 & 312.0$\pm$5.0 & 3 \\
1993.734 & V & 13.18$\pm$0.05 & 13.58$\pm$0.07 & 89.0$\pm$2.0 & 311.2$\pm$1.3 & 4 \\
1993.816 & V & 12.34$\pm$0.04 & 13.52$\pm$0.07 & 93.0$\pm$2.0 & 313.1$\pm$1.3 & 4 \\
1993.901 & K & & & 96.0$\pm$4 & 309.5$\pm$0.7 & 3 \\
1994.567 & UBVRI & & & 87.1$\pm$3.8 & 301.2$\pm$2.0 & 5 \\
1994.797 & K & & & 89.0$\pm$2 & 302.0$\pm$3.0 & 3 \\
1994.821 & V & 13.12$\pm$0.05 & 13.43$\pm$0.06 & 91.2$\pm$2.0 & 302.1$\pm$1.3 & 4 \\
1994.940 & 658 nm & & & 94$\pm$2 & 120.6$\pm$2.5 & 2 \\
1994.966 & K & & & 89.0$\pm$1 & 301.0$\pm$1.0 & 3 \\
1995.055 & V & 12.87$\pm$0.05 & 13.57$\pm$0.06 & 91.6$\pm$2.0 & 302.6$\pm$1.3 & 4 \\
1995.572 & V & 12.57$\pm$0.04 & 13.45$\pm$0.06 & 90.7$\pm$2.0 & 302.3$\pm$1.3 & 4 \\
1996.934 & K & & & 89.6$\pm$7.9 & 290.2$\pm$1.2 & 5 \\
1997.019 & V & 12.82$\pm$0.05 & 13.57$\pm$0.05 & 94.8$\pm$2 & 288.6$\pm$1.3 & 6 \\
1997.706 & V & 13.00$\pm$0.05 & 13.53$\pm$0.07 & 94.9$\pm$2 & 285.4$\pm$1.3 & 6 \\
1997.884 & V & 12.44$\pm$0.05 & 13.55$\pm$0.06 & 96.6$\pm$2 & 284.8$\pm$1.3 & 6 \\
1998.164 & V & 11.99$\pm$0.04 & 13.39$\pm$0.06 & 93.6$\pm$2 & 280.5$\pm$1.3 & 6 \\
1998.775 & R$'$ & 8.9 & 9.1 & 96$\pm$2 & 97.6$\pm$1.2 & 7 \\
1999.695 & V & 12.83$\pm$0.04 & 13.53$\pm$0.05 & 98.5$\pm$2.0 & 272.2$\pm$1.2 & 6 \\
2000.241 & V & 11.77$\pm$0.02 & 13.50$\pm$0.05 & 100.4$\pm$2.0 & 267.5$\pm$1.2 & 6 \\
2000.671 & V & 13.01$\pm$0.04 & 13.53$\pm$0.05 & 100.9$\pm$2.0 & 266.8$\pm$1.2 & 6 \\
2000.695\tablenotemark{c} & V & 12.74$\pm$0.04 & 13.57$\pm$0.05 & & & 6 \\
2001.063 & V & 12.60$\pm$0.03 & 13.52$\pm$0.05 & 100.5$\pm$2.0 & 267.9$\pm$1.2 & 6 \\
2001.164 & V & 13.12$\pm$0.04 & 13.42$\pm$0.05 & 100.5$\pm$2.0 & 265.1$\pm$1.2 & 6 \\
2002.129 & V & 12.68$\pm$0.03 & 13.51$\pm$0.05 & 102.1$\pm$2.0 & 262.3$\pm$1.2 & 6 \\
\cutinhead{ZZ Tau (HBC 46), B-V = 1.5\tablenotemark{b} mag}
1994.934\tablenotemark{d} & V & 14.7$\pm$0.1 & 15.9$\pm$0.2 &  &  & 4 \\
1995.081\tablenotemark{e} &   & & & 36.0$\pm$7.6 & 187.0$\pm$11 & \\
1995.227\tablenotemark{d} & V & 14.9$\pm$0.1 & 15.7$\pm$0.2 &  &  & 4 \\
1995.994\tablenotemark{d} & V & 15.0$\pm$0.1 & 15.5$\pm$0.2 &  &  & 6 \\
1996.124\tablenotemark{e} &   & & & 36.0$\pm$7.6 & 177.0$\pm$10 & \\
1996.145\tablenotemark{d} & V & 15.0$\pm$0.1 & 15.5$\pm$0.2 & &  & 4 \\
1996.235\tablenotemark{d} & V & 15.0$\pm$0.1 & 15.5$\pm$0.2 &  &  & 6 \\
1997.019\tablenotemark{d} & V & 14.8$\pm$0.1 & 15.9$\pm$0.2 &  &  & 6 \\
1997.704 & V & 15.0$\pm$0.1 & 15.6$\pm$0.2 & 50.6$\pm$5.4 & 129.2$\pm$6.0 & 6 \\
1998.214 & V & 14.9$\pm$0.1 & 15.8$\pm$0.2 & 55.0$\pm$5.4 & 125.0$\pm$6.0 & 6 \\
1999.621 & V & 14.5$\pm$0.1 & 15.9$\pm$0.2 & 57.9$\pm$1.3 & 112.9$\pm$1.8 & 6 \\
2000.564 & V & 14.8$\pm$0.1 & 15.9$\pm$0.2  & 59.5$\pm$1.2 & 106.2$\pm$1.8 & 6 \\
2000.690 & V & 14.8$\pm$0.1 & 15.9$\pm$0.2 & 58.2$\pm$1.2 & 105.7$\pm$1.8 & 6 \\
2001.646 & V & 14.6$\pm$0.1 & 16.0$\pm$0.2 & 62.5$\pm$1.1 & 98.9$\pm$1.8 & 6 \\
2002.748 & V & 14.7$\pm$0.1 & 15.9$\pm$0.2 & 60.6$\pm$1.3 & 91.1$\pm$1.7 & 6 \\
2002.929 & V & 14.7$\pm$0.1 & 16.0$\pm$0.2 & 61.2$\pm$1.5 & 88.8$\pm$1.5 & 6 \\
\cutinhead{Elias 12 (HBC 404), (B-V)$_S$ = 1.18\tablenotemark{a} mag, (B-V)$_N$ = 1.24\tablenotemark{a} mag}
1990.857 & K & & & 371.0$\pm$0.5 & 332.5$\pm$0.2 & 3 \\
1993.901 & K & & & 340$\pm$20 & 334$\pm$6 & 3 \\
1994.728 & K & & & 336$\pm$3 & 326.9$\pm$0.9 & 3 \\
1994.794 & K & & & 329$\pm$4 & 328.1$\pm$0.6 & 3 \\
1994.654 & V & 12.07$\pm$0.03 & 13.83$\pm$0.14 & 335.4$\pm$2.0 & 328.5$\pm$0.4 & 4 \\
1995.096 & V & 12.77$\pm$0.04 & 14.47$\pm$0.08 & 337.5$\pm$2.0 & 328.8$\pm$0.4 & 4 \\
1997.923 & K & 7.36$\pm$0.05 & 8.21$\pm$0.10 & 308.9$\pm$5.4 & 323.2$\pm$1.0 & 5 \\
1997.931 & L & 6.54$\pm$0.14 & 7.94$\pm$0.14 & 314.9$\pm$5.6 & 323.3$\pm$1.0 & 5 \\
1998.925 & UBVRI & & & 299.2$\pm$2.0 & 322.87$\pm$0.14 & 5 \\
1999.755 & V & & & & & 6 \\
S-Na     &  & 12.36$\pm$0.03 & 14.9$\pm$0.4 & 295.8$\pm$4.9 & 325.5$\pm$1.2 & \\
S-Nb	 &  & & 14.9$\pm$0.4 & 294.0$\pm$5.0 & 320.8$\pm$1.2 & \\
Na-Nb 	 &  & & & 24.2$\pm$8.8 & 228.8$\pm$17.1 & \\
2000.662 & V & & & & & 6 \\
S-Na     &  & 12.34$\pm$0.03 & 14.7$\pm$0.2 & 292.5$\pm$3.5 & 324.1$\pm$0.7 & \\
S-Nb     &  & & 15.1$\pm$0.4 & 288.9$\pm$3.9 & 317.9$\pm$0.8 & \\
Na-Nb    &  & & & 31.5$\pm$5.6 & 224.3$\pm$9.5 &  \\
2001.001 & V & & & & & 6 \\
S-Na	 &  & 12.36$\pm$0.03 & 14.7$\pm$0.2 & 285.3$\pm$5.3 & 323.8$\pm$1.0 & \\
S-Nb	 &  & & 15.3$\pm$0.4 & 278.6$\pm$7.3 & 317.0$\pm$1.5 & \\
Na-Nb    &  & & & 34.0$\pm$9.0 & 219.2$\pm$14.9 & \\
2001.170 & V & & & & & 6 \\
S-Na	 &  & 12.34$\pm$0.03 & 14.7$\pm$0.3 & 284.7$\pm$5.2 & 323.5$\pm$1.2 & \\
S-Nb	 &  & & 15.3$\pm$0.4 & 272.6$\pm$7.7 & 317.4$\pm$1.5 & \\
Na-Nb    &  & & & 32.1$\pm$9.3 & 208.3$\pm$16.3 & \\
2002.182 & H K$'$ & & & & & 8 \\
S-Na	 &  & & & 284.3$\pm$2.0 & 320.7$\pm$0.6 & \\   
S-Nb	 &  & & & 257.4$\pm$1.8 & 314.1$\pm$0.4 & \\   
Na-NB    &  & & & 42.6$\pm$3.2 & 188.3$\pm$5.8 & \\   
2002.828 & H K$'$ & & & & & 8 \\
S-Na	 &  & & & 283.5$\pm$2.0 & 319.0$\pm$0.5 & \\  
S-Nb	 &  & & & 249.0$\pm$1.8 & 313.0$\pm$0.4 & \\  
Na-Nb    &  & & & 44.3$\pm$2.8 & 175.0$\pm$4.6 & \\  
2002.972 & V & & & & & 6 \\
S-Na	 &  & 12.36$\pm$0.03 & 14.6$\pm$0.2 & 280.3$\pm$4.9 & 319.8$\pm$0.8 & \\
S-Nb	 &  & & 15.2$\pm$0.4 & 245.0$\pm$6.4 & 313.3$\pm$1.5 & \\
Na-Nb    &  & & & 46.0$\pm$7.3 & 176.6$\pm$10.2 & \\
\enddata 
\tablenotetext{a}{White \& Ghez 2001}
\tablenotetext{b}{Observed as an unresolved system; Herbig \& Bell 1988}
\tablenotetext{c}{Magnitude from the x-axis solution; y-axis solution unreliable.}
\tablenotetext{d}{Magnitude from the y-axis solution; x-axis solution unreliable.}
\tablenotetext{e}{Values interpolated from Simon et al. 1996; see text.}
\tablerefs{(1) Chen 1990; (2) Thi\'{e}baut et al. 1995; (3) Ghez et al. 1995; (4) Simon, Holfeltz \& Taff 1996; (5) White \& Ghez 2001; (6) This work - FGS (7)  Balega et al. 2002; (8) This work - NIRC2 AO.}
\end{deluxetable} 

\clearpage
 
\begin{deluxetable}{lcll} 
\tablewidth{0pt}
\tablecaption{Flux Ratios of Elias 12} 
\tablehead{\colhead{Date} & \colhead{Filter} & \colhead{S : Na : Nb} & \colhead{Nb/Na}}
\startdata 
1999.755 & V & 1.00 : 0.10$\pm$0.03 : 0.09$\pm$0.03 & 0.98$\pm$0.50 \\
2000.662 & V & 1.00 : 0.11$\pm$0.02 : 0.08$\pm$0.03 & 0.72$\pm$0.27 \\
2001.001 & V & 1.00 : 0.11$\pm$0.02 : 0.07$\pm$0.02 & 0.61$\pm$0.23 \\
2001.170 & V & 1.00 : 0.11$\pm$0.02 : 0.07$\pm$0.02 & 0.61$\pm$0.28 \\
2002.182 & H & 1.00 : 0.38$\pm$0.01 : 0.29$\pm$0.01 & 0.76$\pm$0.04 \\   
2002.182 & K$'$ & 1.00 : 0.39$\pm$0.01 : 0.28$\pm$0.01 & 0.72$\pm$0.02 \\   
2002.828 & H & 1.00 : 0.37$\pm$0.01 : 0.27$\pm$0.01 & 0.73$\pm$0.04 \\  
2002.828 & K$'$ & 1.00 : 0.31$\pm$0.01 : 0.23$\pm$0.01 & 0.75$\pm$0.03 \\  
2002.972 & V & 1.00 : 0.13$\pm$0.02 : 0.08$\pm$0.02 & 0.58$\pm$0.20 \\
\enddata 
\end{deluxetable} 

\clearpage

\begin{figure}
	\scalebox{0.4}{\includegraphics{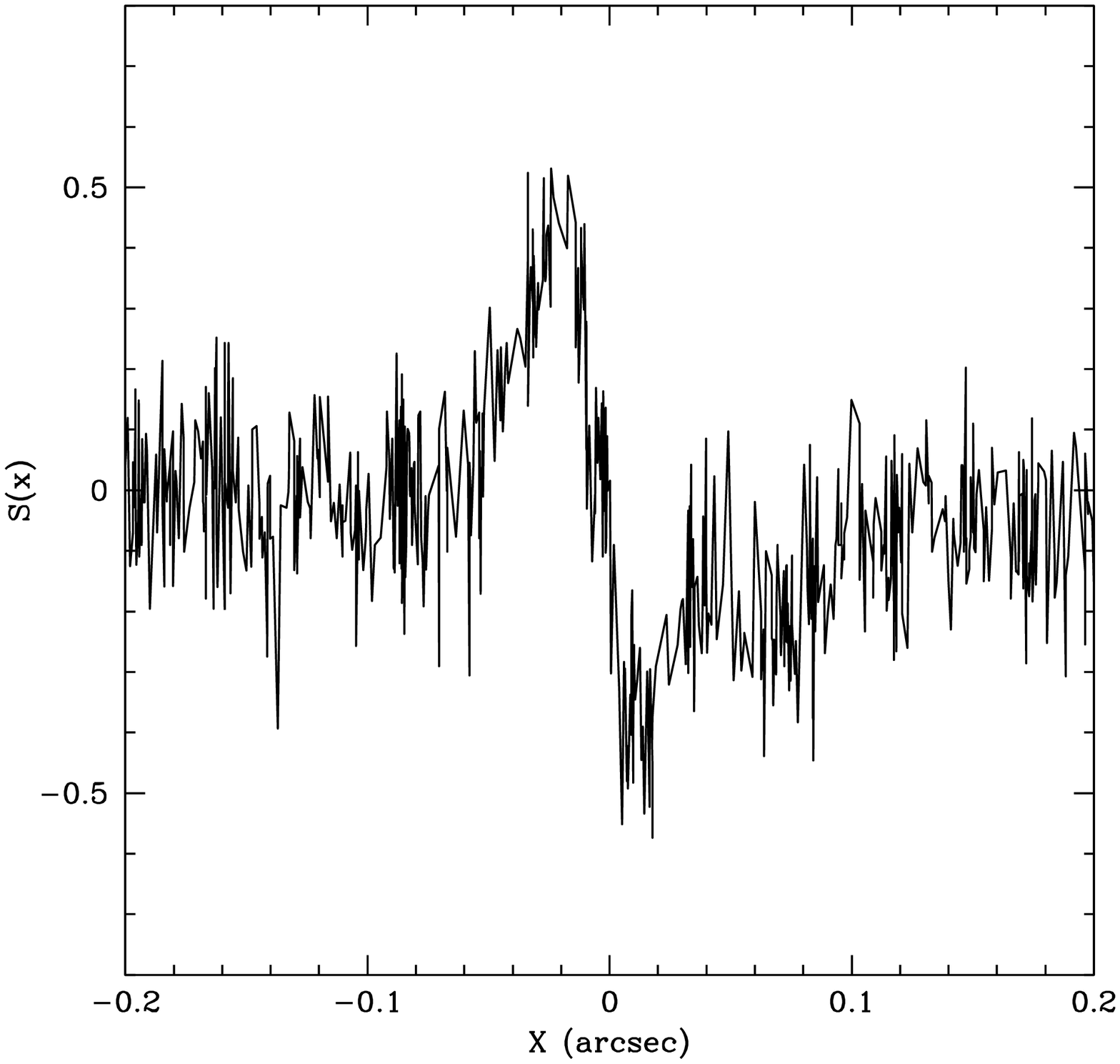}}
	\scalebox{0.4}{\includegraphics{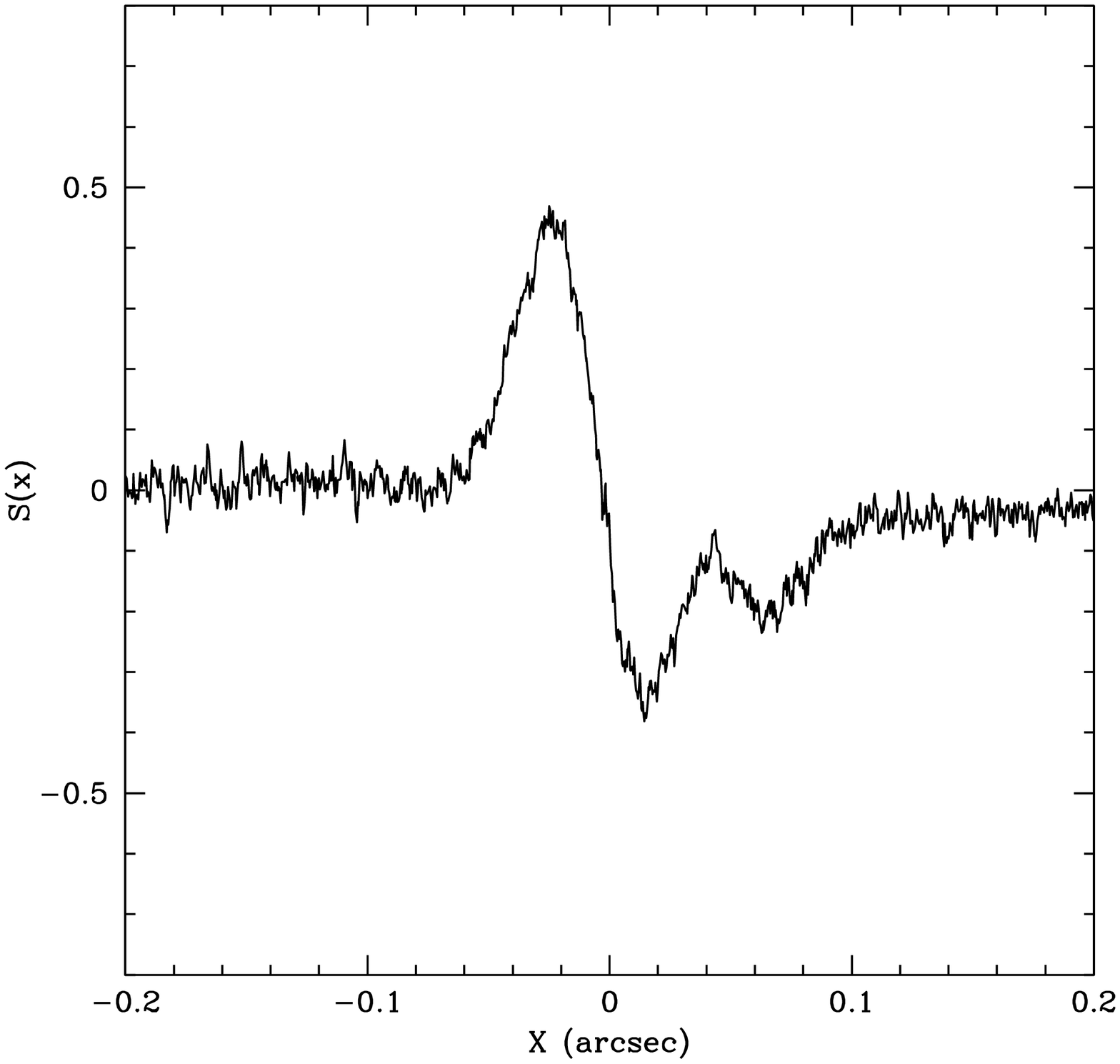}} \\
	\scalebox{0.4}{\includegraphics{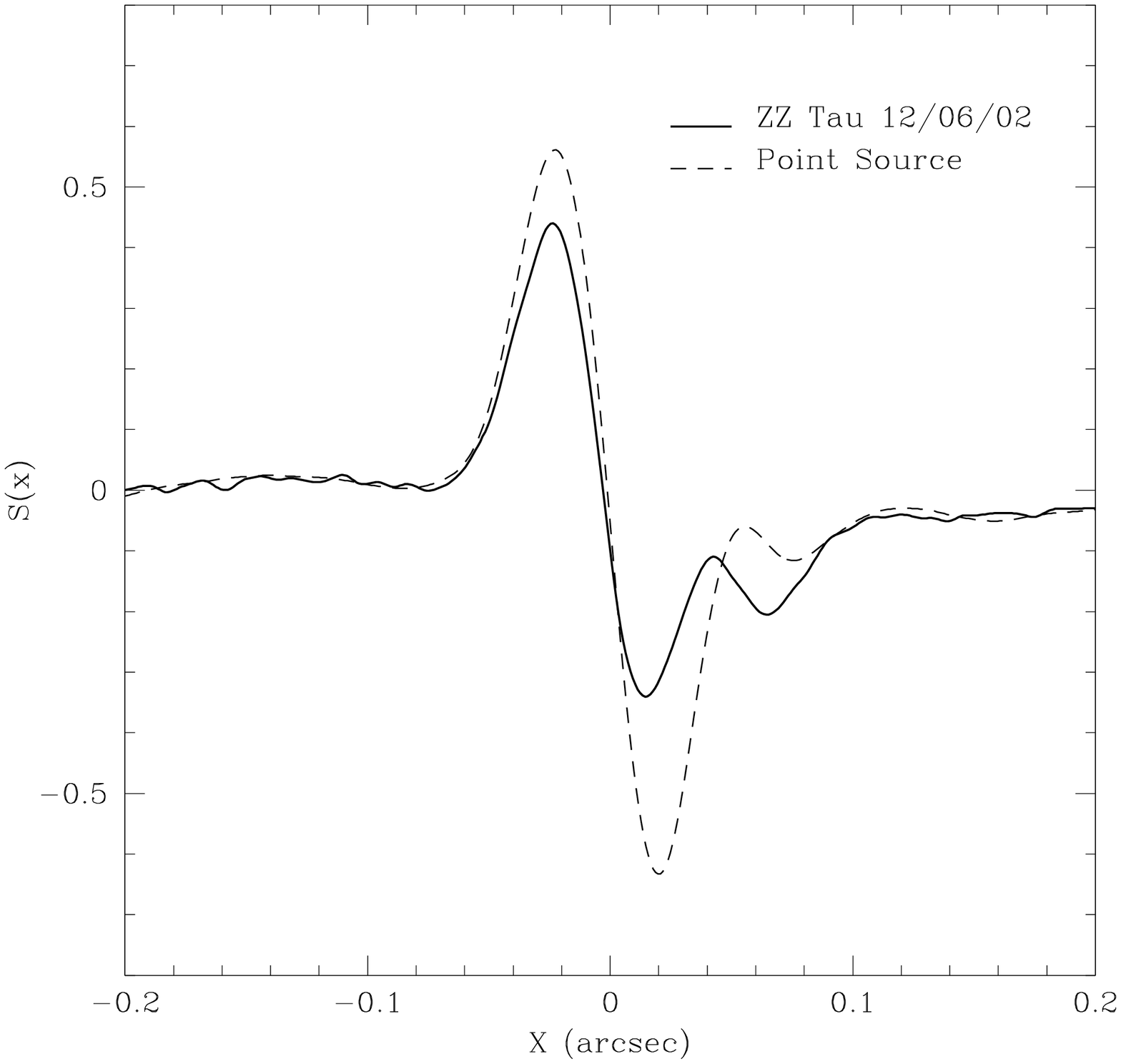}}
	\scalebox{0.4}{\includegraphics{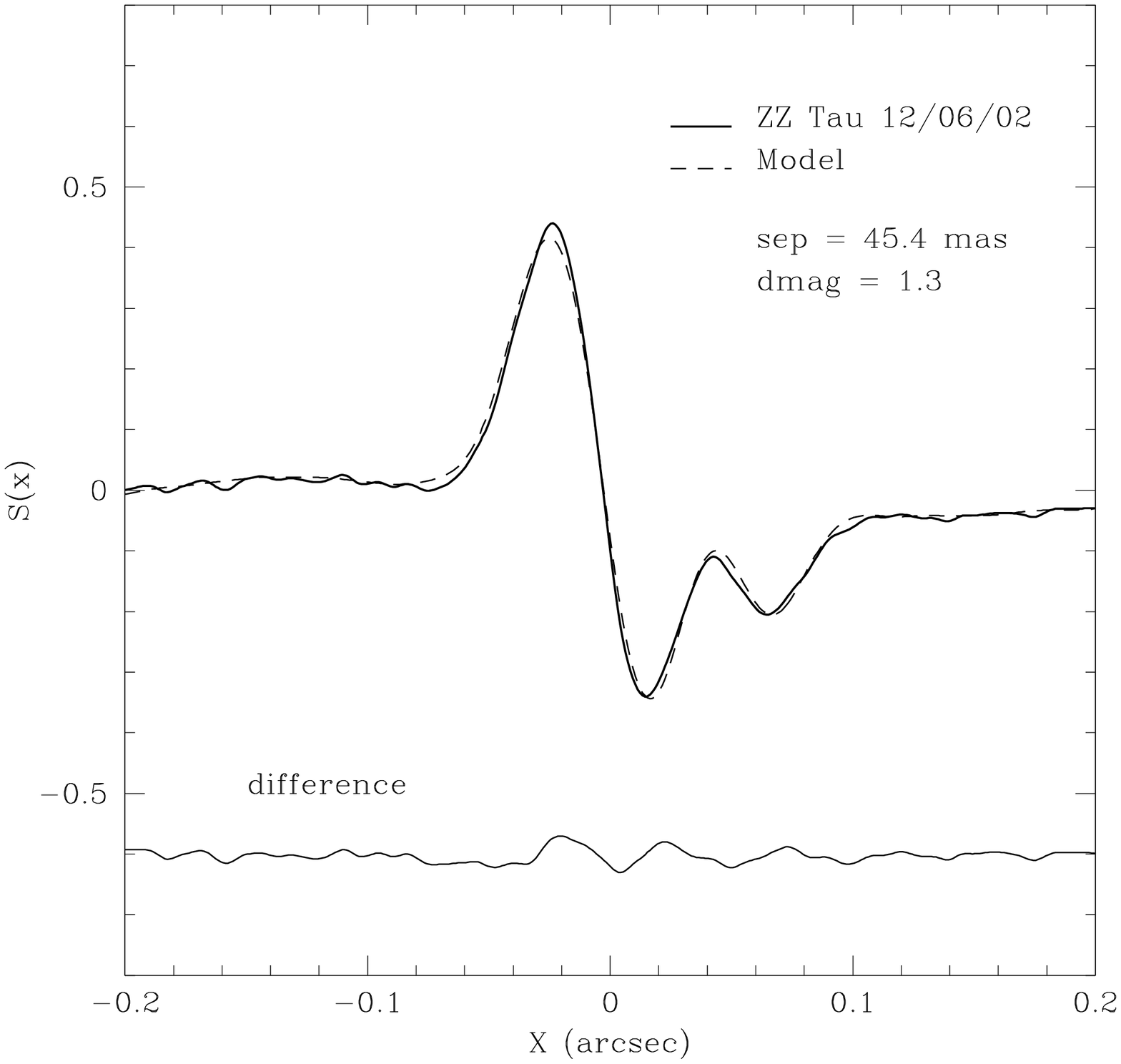}}
	\caption{(a) Single raw FGS scan of ZZ Tau along the x-axis, from 6 December 2002. (b) Cross-correlated, co-added product from 20 scans along the x-axis.  (c) Smoothed scan compared to that from a point source (SAO 185689).  (d) Fit of the model solution compared with the data, and the difference between the two.}
\end{figure}

\clearpage

\begin{figure}
	\scalebox{0.54}{\includegraphics{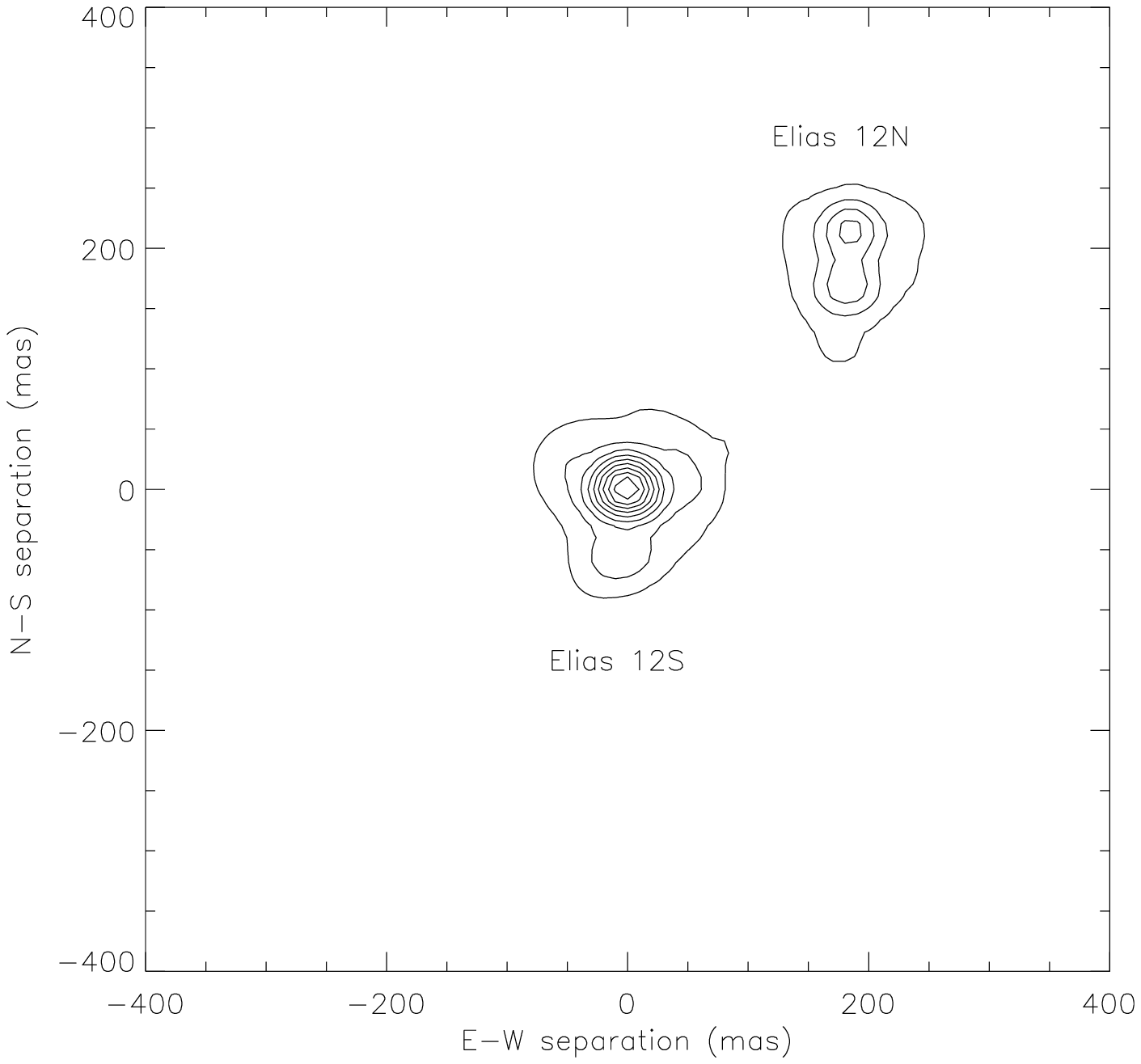}}
	\scalebox{0.54}{\includegraphics{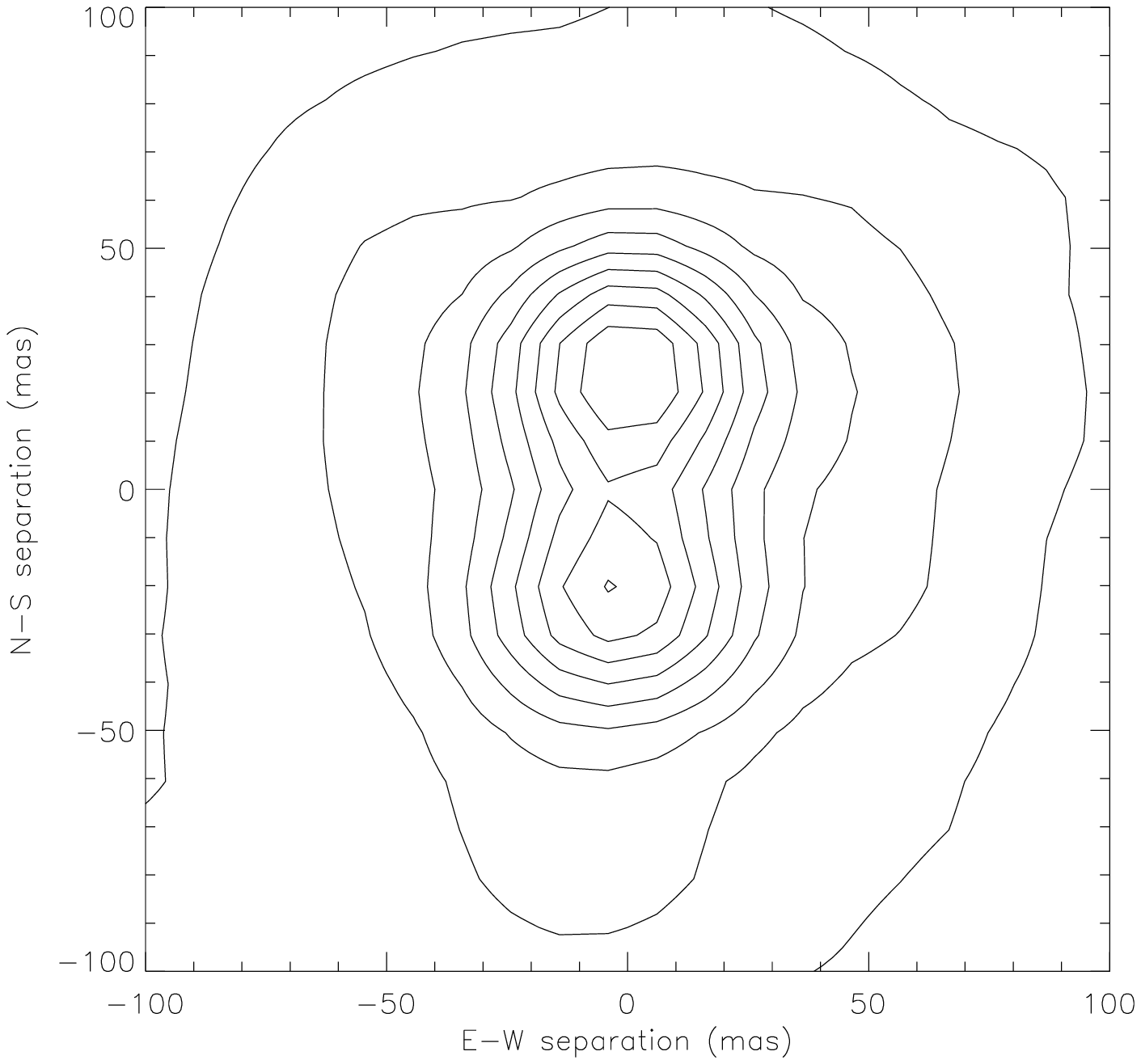}} \\
	\scalebox{0.54}{\includegraphics{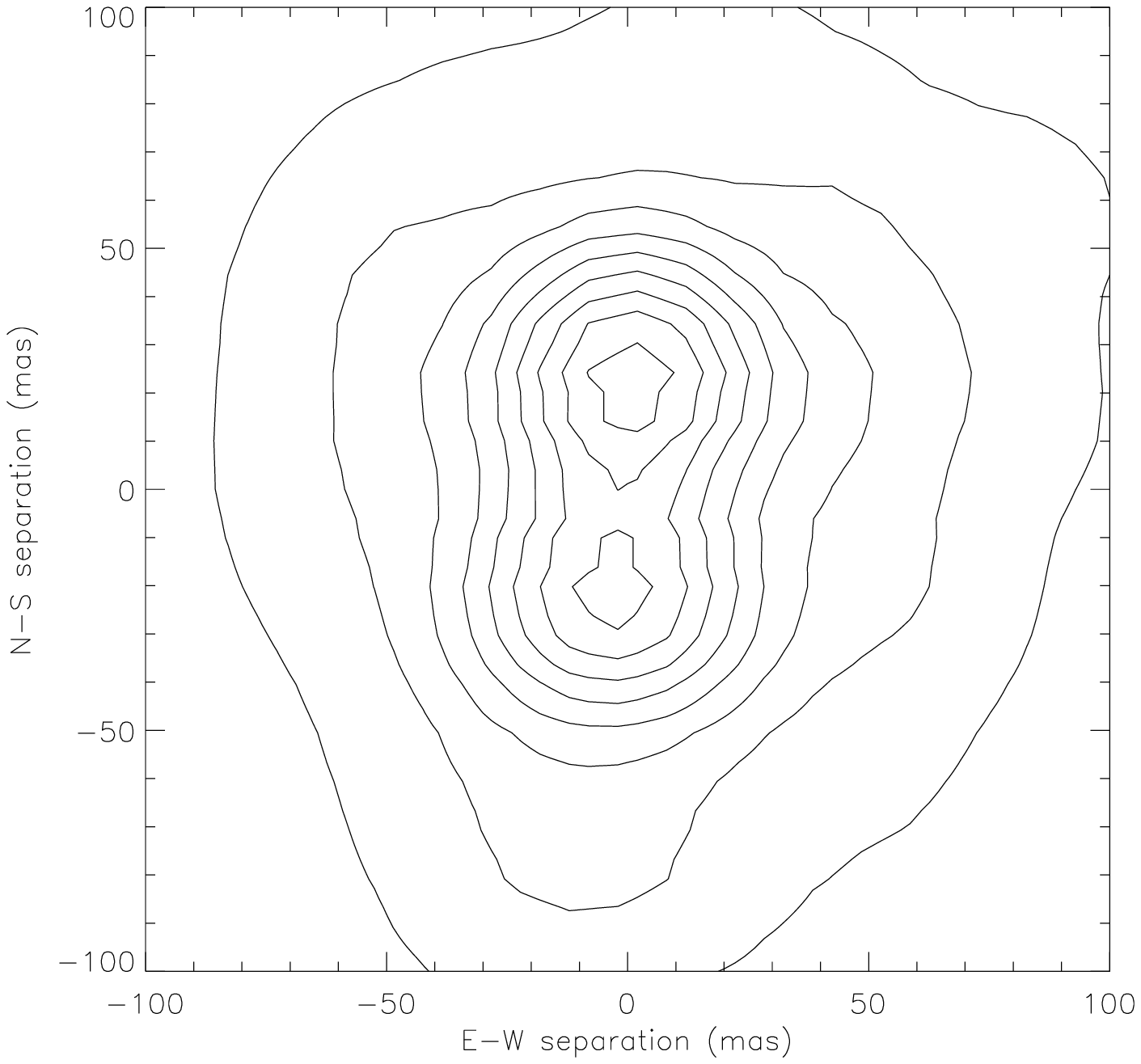}}
	\scalebox{0.54}{\includegraphics{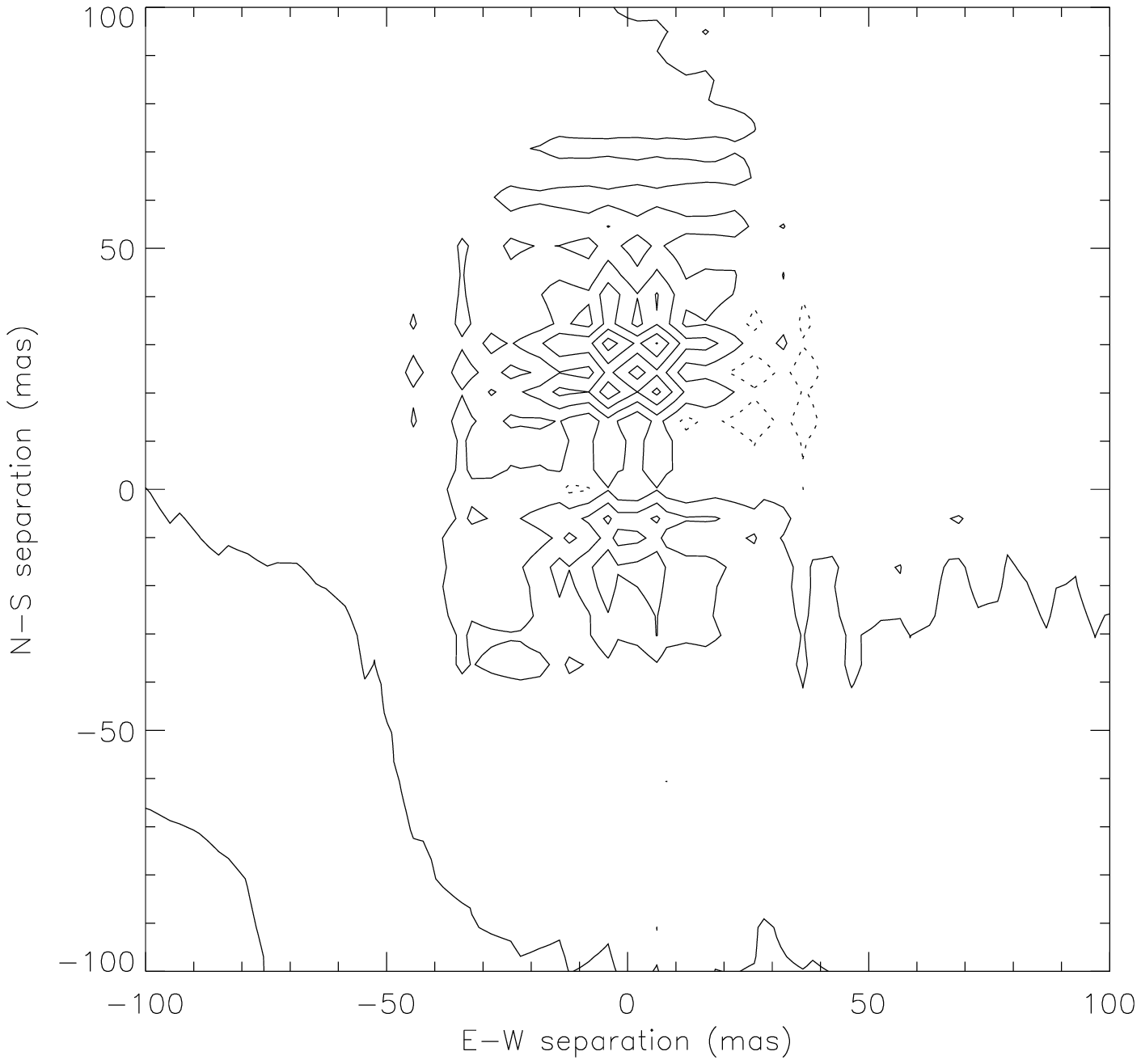}}
   \caption{(a) NIRC-2 image of Elias 12 in the H-band, from 30 October 2002. The image has been re-binned to expand the pixel scale.  The contour levels are spaced evenly at intervals of 10\% of the peak flux in El 12S.  (b) Image of El 12N, centered on the close pair.  The contours go from 10\% to 90\% of the peak flux in the close pair at 10\% intervals.  (c) Model of the close pair El 12N, using El 12S as the PSF.  The contour spacing is the same as in (b).  (d) Difference between the image and the model of the close pair.  Contour levels are spaced at 2.5\% intervals from -5\% to 10\% of the peak flux in the image of the close pair (b).  The negative contour levels are marked by dotted lines. }
\end{figure}

\clearpage

\begin{figure}
   \includegraphics{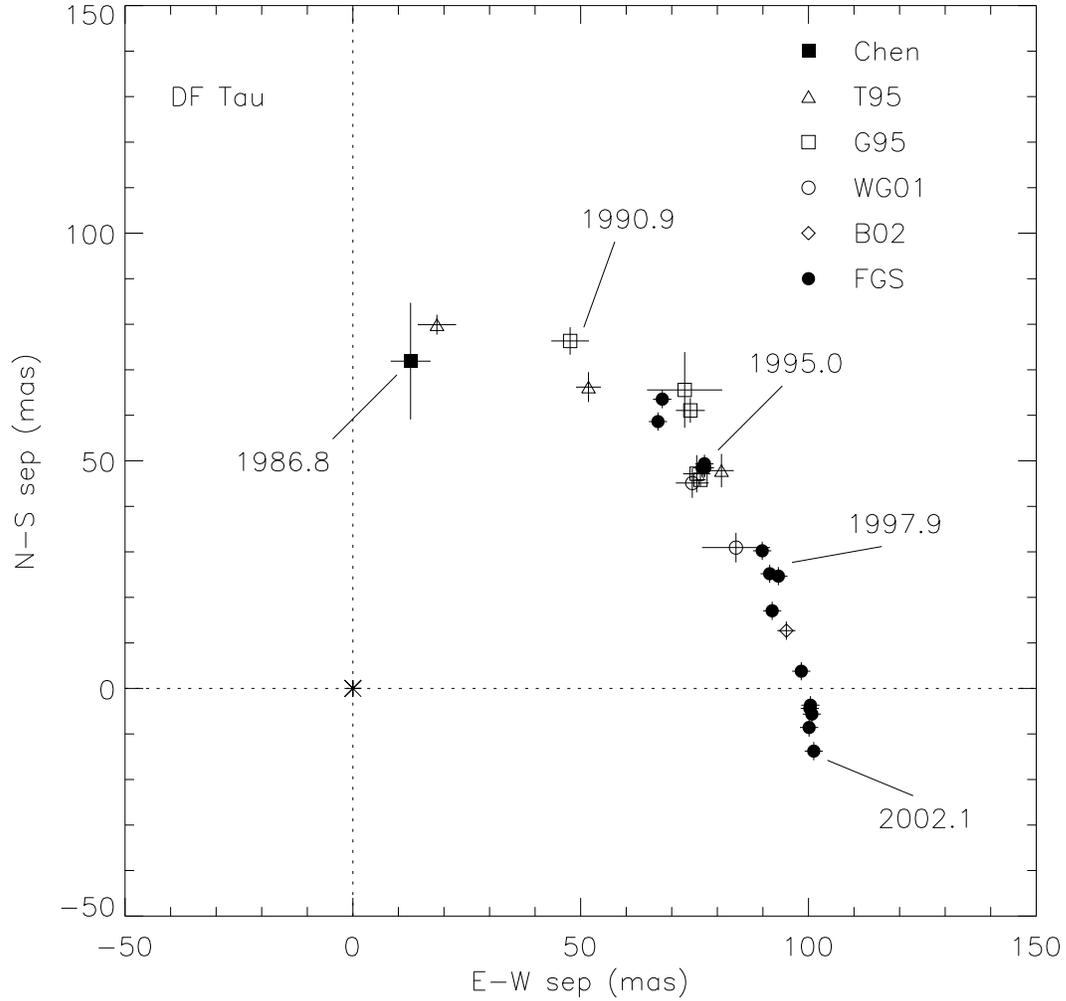}
   \caption{Position of the secondary relative to the primary for DF Tau.  The location of the primary star is marked by an asterisk. The data points for DF Tau were measured by Chen (1990), Thiebaut et al. (1995, T95), Ghez et al. (1995, G95), White \& Ghez (2001, WG01), Balega et al. (2002, B02), and our observations with the FGS.}
\end{figure}

\clearpage

\begin{figure}
   \includegraphics{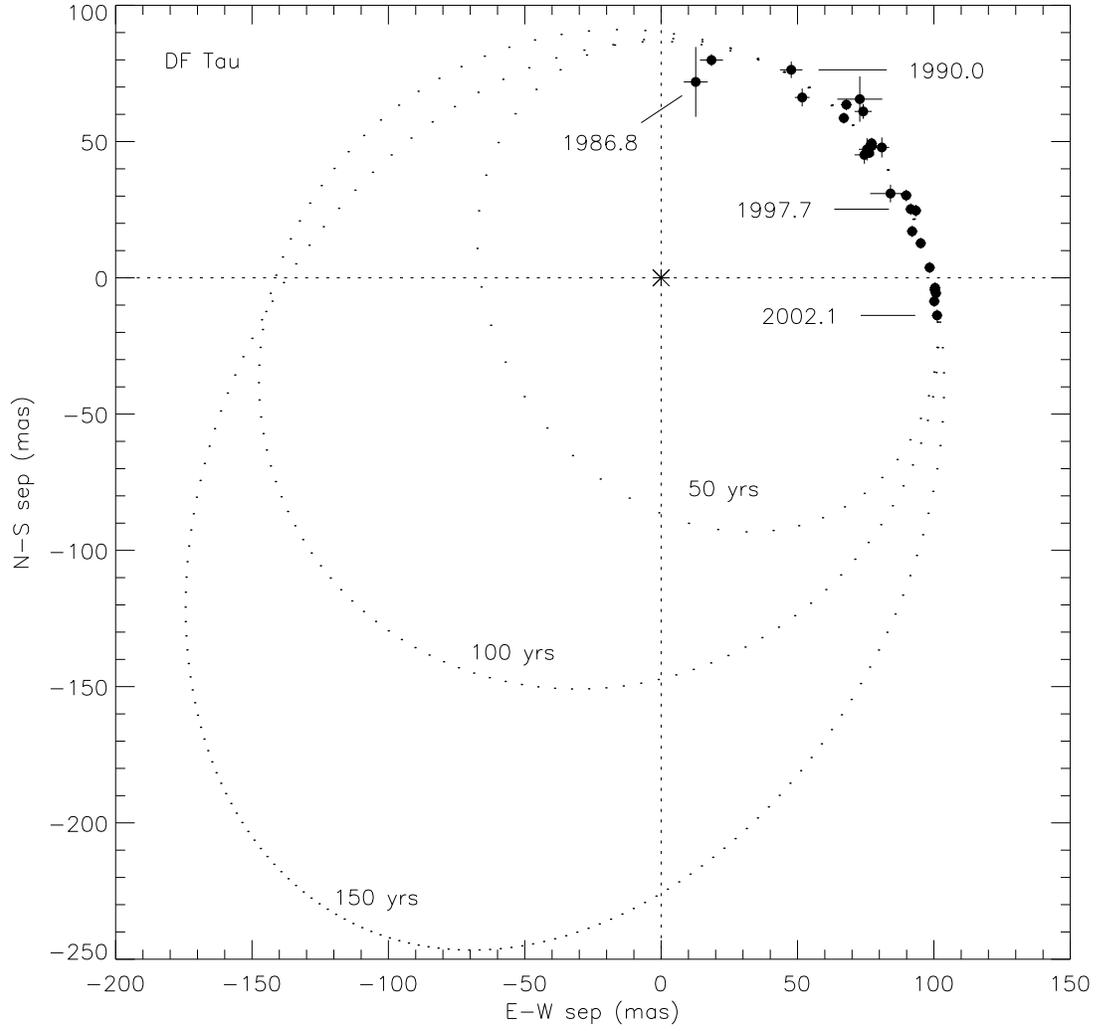}
   \caption{Examples of three possible orbital solutions for DF Tau at periods of 50, 100, and 150 years.  The calculated orbits are plotted at yearly intervals.}
\end{figure}

\clearpage

\begin{figure}
   \includegraphics{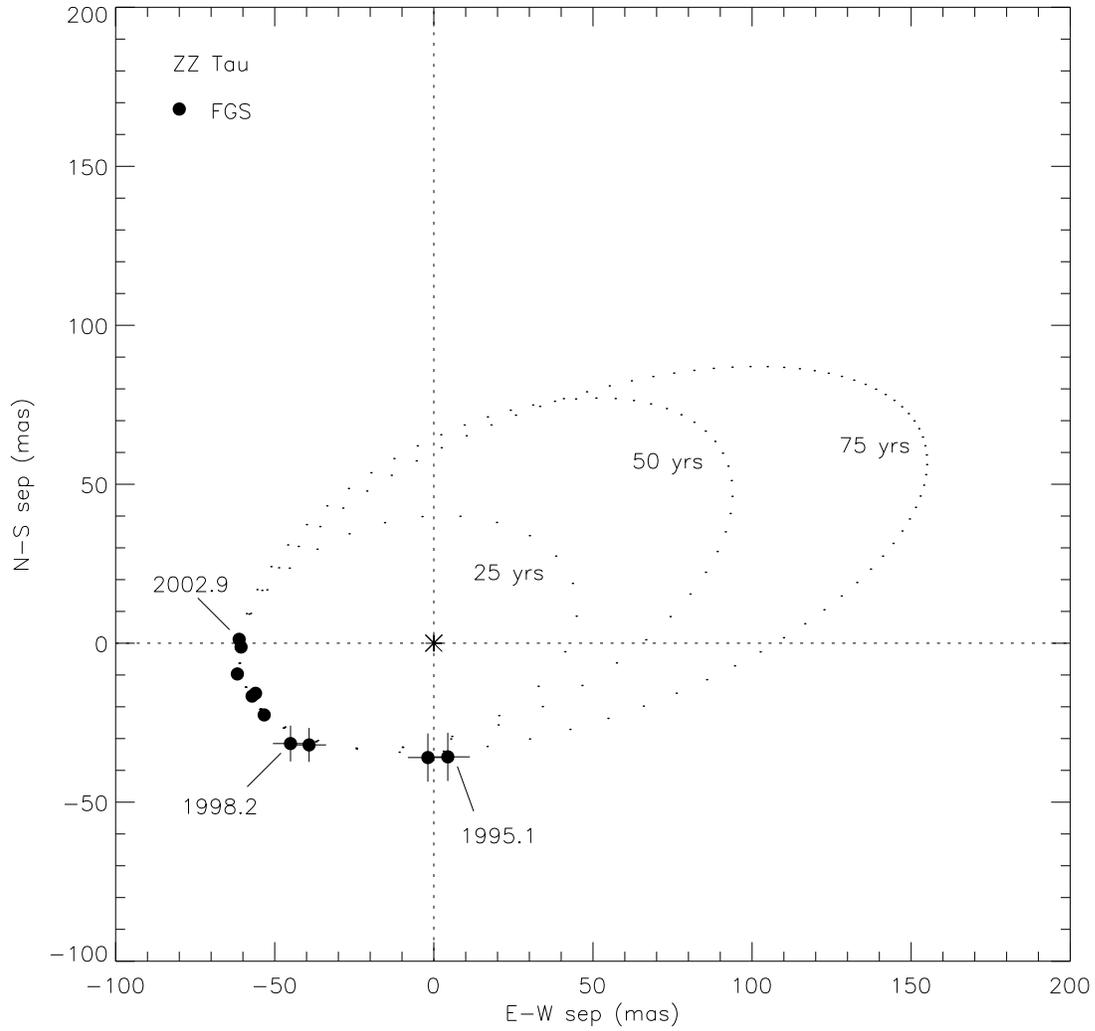}
   \caption{Position of the secondary relative to the primary for ZZ Tau, as measured by the FGS.  The location of the primary star is marked by an asterisk.  Examples of three possible orbital solutions for ZZ Tau, at periods of 25, 50, and 75 years, are plotted at yearly intervals.}
\end{figure}

\clearpage

\begin{figure}
   \includegraphics{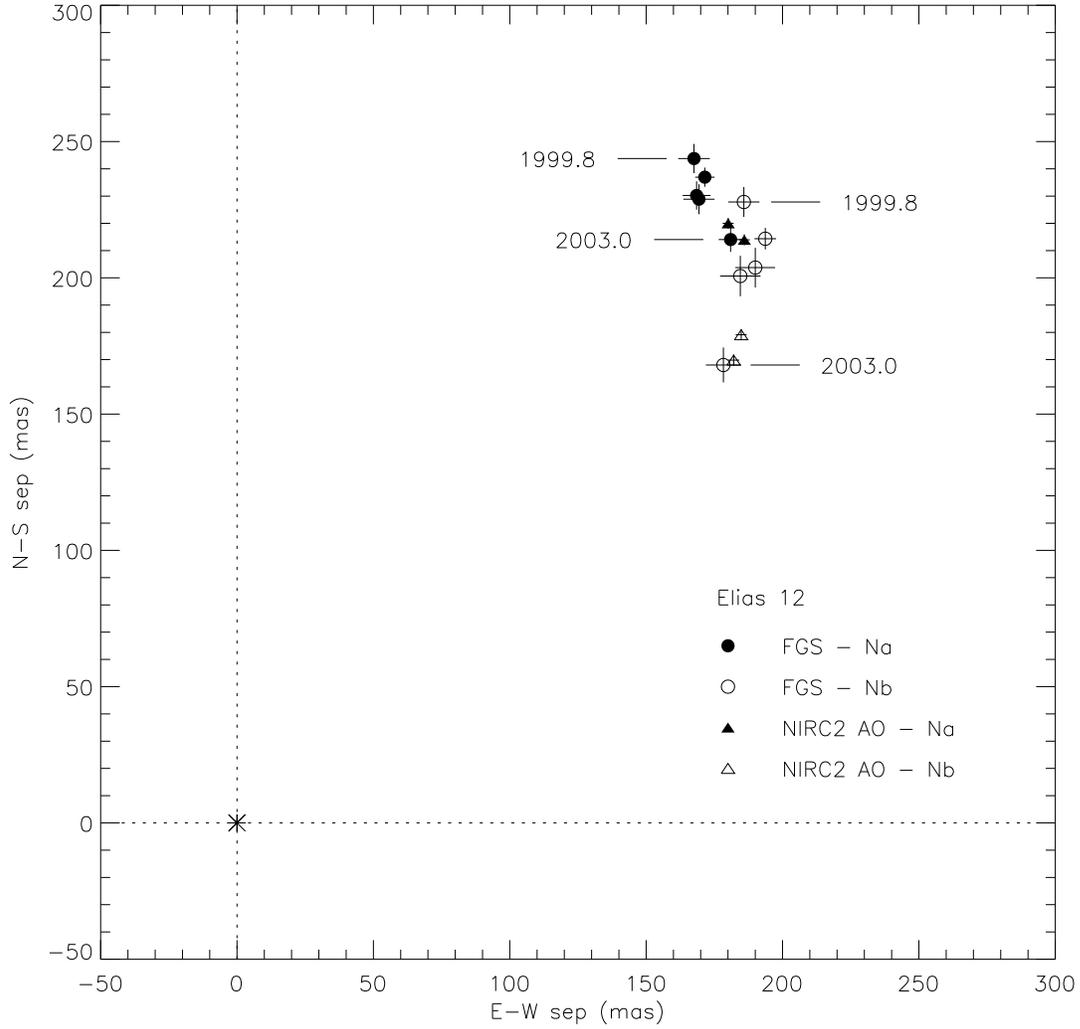}
   \caption{Orbital motion in the Elias 12 hierarchical triple.  The positions of Elias 12 Na (filled symbols) and Nb (open symbols) are plotted relative to Elias 12S (marked by an asterisk).  The measurements are from our observations with the FGS and NIRC-2 with adaptive optics.}
\end{figure}

\clearpage

\begin{figure}
   \includegraphics{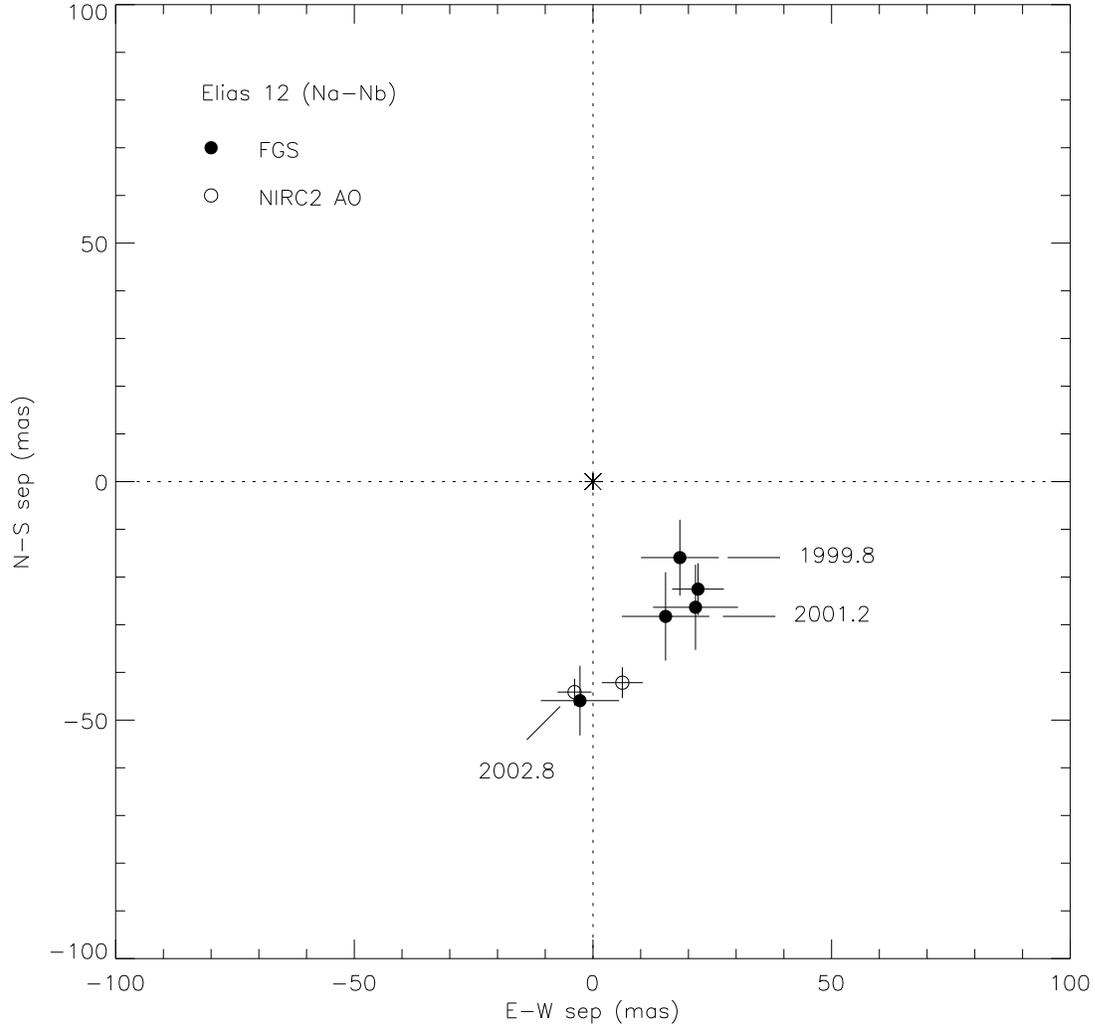}
   \caption{Orbital motion of the close pair in Elias 12, Na-Nb (see text and Figure 6).  The motion of the fainter component, Nb, is plotted relative to Na (marked by an asterisk).}
\end{figure}

\clearpage

\begin{figure}
   \scalebox{0.9}{\includegraphics{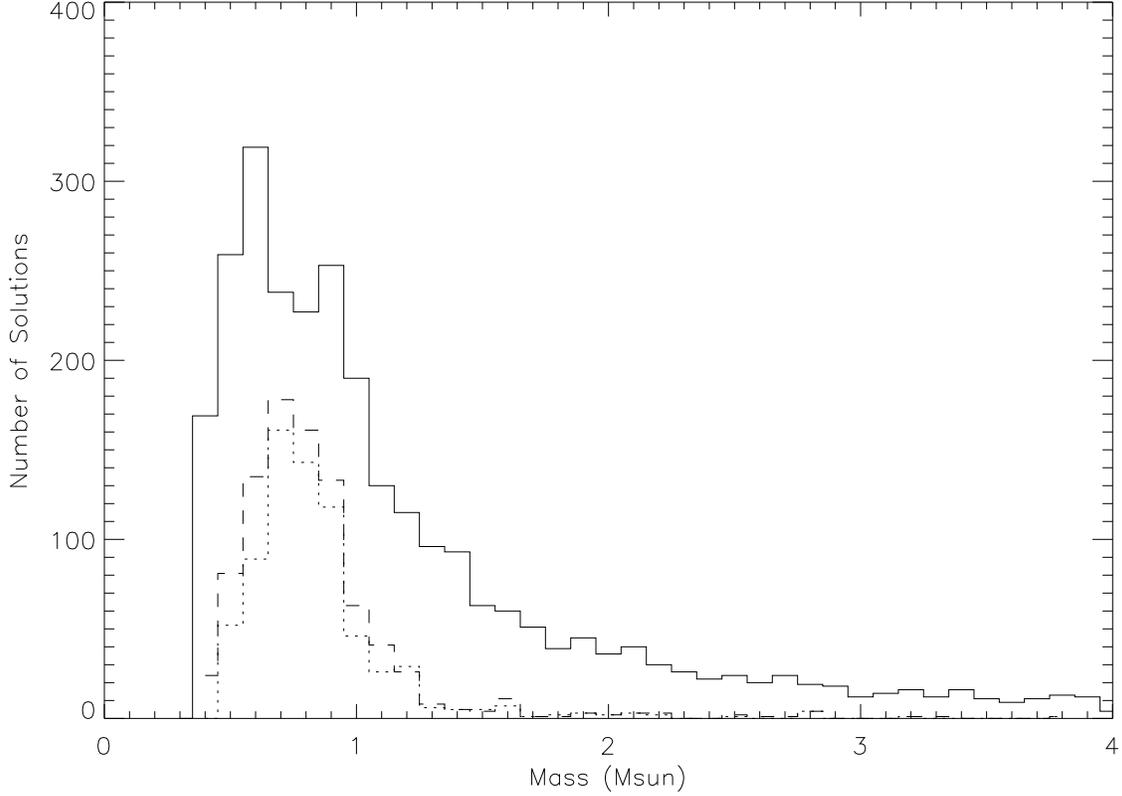}}
   \caption{Distribution of masses for DF Tau.  The solid line shows the mass distribution allowed by the current measurements to date (see text).  The dotted line shows the mass distribution resulting from simulated annual observations for another 5 years into the future from 2003.0, assuming a 50 year period.  The dashed line shows the mass distribution resulting from simulated annual observations for another 5 years into the future from 2003.0, assuming a 100 year period.  All three of the mass distributions are calculated on the same grid; the smaller numbers in the simulated distributions inidicate the elimination of possible orbits.  The values of the total mass for the three distributions are (0.90$^{+0.85}_{-0.35}$), (0.79$^{+0.21}_{-0.15}$), and (0.77$^{+0.23}_{-0.15}$) M$_\sun$, at a distance of 140 pc.}
\end{figure}

\clearpage

\begin{figure}
   \scalebox{0.9}{\includegraphics{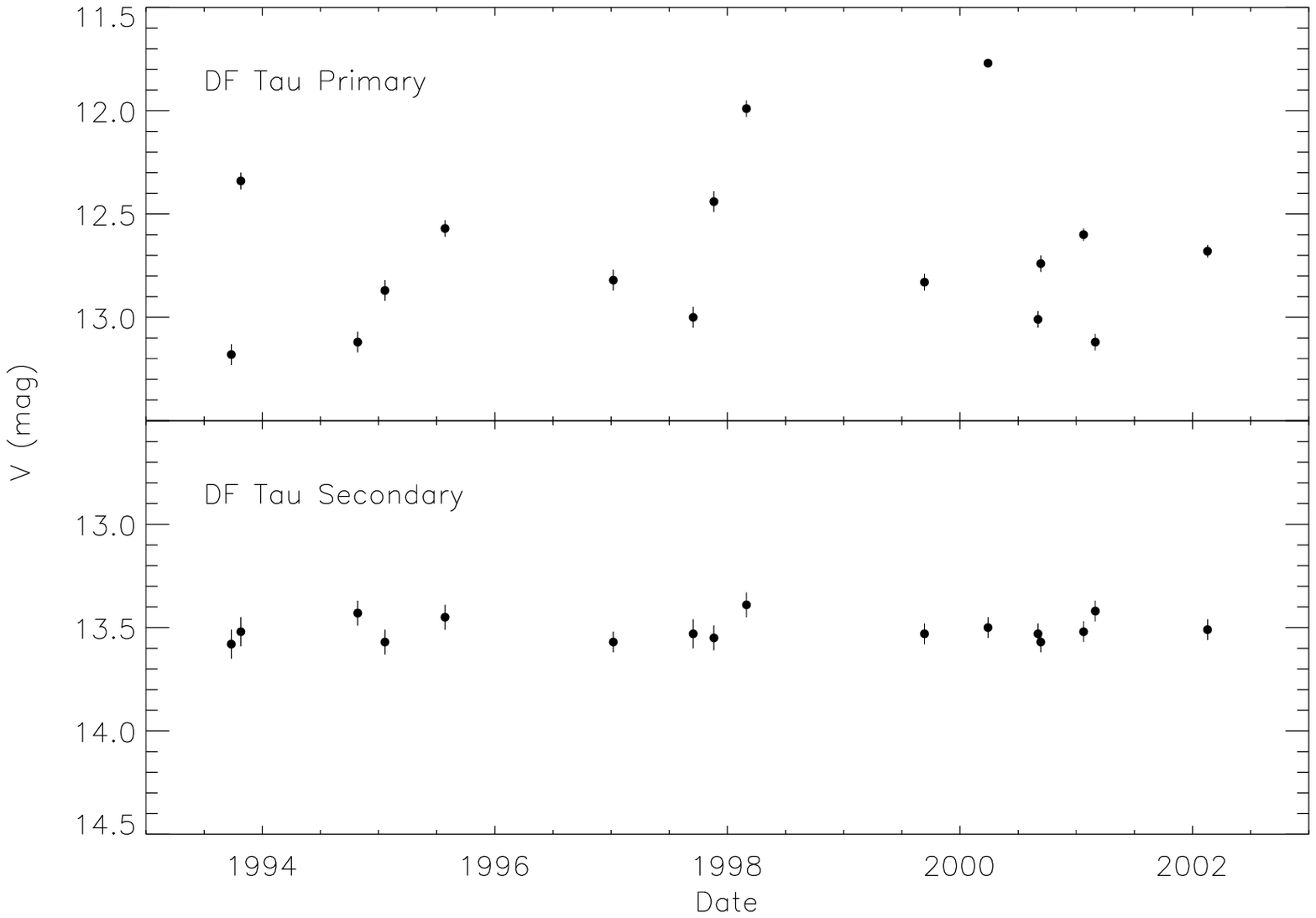}}
   \caption{Photometric variability of DF Tau.  The upper panel shows the magnitude of the primary component during each FGS observation.  The lower panel shows that the variability of the DF Tau binary is mostly attributable to the primary.}
\end{figure}


\begin{thebibliography}{}
\bibitem[Aitken 1964]{aitken} Aitken, R. G.  1964, The Binary Stars (NY: Dover Publications)
\bibitem[Balega et al. 2002]{bal02} Balega, I. I., Balega, Y. Y., Hofmann, K. H., Maksimov, A. F., Pluzhnik, E. A., Schertl, D., Shkhagosheva, Z. U., \& Weigelt, G. 2002, A\&A, 385, 87
\bibitem[Baraffe et al. 1998]{bcah98} Baraffe, I., Chabrier, G., Allard, F., \& Hauschildt, P. H. 1998, A\&A, 337, 403
\bibitem[Bevington \& Robinson 1992]{bev92} Bevington, P. R. \& Robinson, D. K. 1992, Data Reduction ans Error Analysis for the Physical Sciences (2nd ed.; NY: McGraw-Hill, Inc.) 
\bibitem[Bouvier \& Bertout 1989]{bouv89} Bouvier, J. \& Bertout, C. 1989, A\&A, 211, 99
\bibitem[Bouvier et al. 1993]{bouv93} Bouvier, J., Cabrit, S., Fern\'{a}ndez, M., Mart\'{i}n, E. L., \& Matthews, J. M. 1993, A\&A, 272, 176
\bibitem[Chen 1990]{chen90} Chen, W. P. 1990, Ph.D. Thesis, State University of New York at Stony Brook
\bibitem[Couteau 1981]{couteau} Couteau, P. 1981,  Observing Visual Double Stars (Cambridge, MA: The MIT Press)
\bibitem[D'Antona \& Mazzitelli 1997]{DM97} D'Antona, F. \& Mazzitelli, I. 1997, Mem. Soc. Astron. Italiana, 68, 807
\bibitem[Eggen 1967]{egg67} Eggen O. J. 1967, ARA\&A, 5, 105
\bibitem[Ghez et al. 1995]{ghez95} Ghez, A. M., Weinberger, A. J., Neugebauer, G., Matthews, K. \& McCarthy, D. W. 1995, AJ, 110, 753
\bibitem[Haralick 1992]{har} Haralick, R. M. 1992,  Computer and Robot Vision (MA: Addison-Wesley Pub. Co.)
\bibitem[Hartigan \& Kenyon 2003]{hk03} Hartigan, P. \& Kenyon, S. J. 2003, ApJ, 583, 334
\bibitem[Hartkopf, McAlister, \& Franz 1989]{hart89} Hartkopf, W. I., McAlister, H. A., \& Franz, O. G. 1989, AJ, 98, 1014
\bibitem[Hartkopf, Mason, \& Worley 2001]{grading} Hartkopf, W. I., Mason, B. D., \& Worley, C. E. 2001, AJ, 122, 3472
\bibitem[Hartkopf \& Mason 2002]{hart02} Hartkopf, W. I. \& Mason, B. D. 2002, Sixth Catalog of Orbits of Visual Binary Stars (http://ad.usno.navy.mil/wds/orb6.html)
\bibitem[Herbig \& Bell 1988]{HBC} Herbig, G. H. \& Bell, K. R. 1988, Lick Obs. Bull., 1111
\bibitem[Herbst et al. 1994]{herbst94} Herbst, W., Herbst, D. K., Grossman, E. J., \& Weinstein, D. 1994, AJ, 108, 1906 (http://www.astro.wesleyan.edu/~bill/research/ttauri.html\#ov2)
\bibitem[Holfeltz et al. 1995]{hol95} Holfeltz, S. T., Nelan, E. P., Taff, L. G., \& Lattanzi, M. G. 1995, {\it Hubble Space Telescope} Fine Guidance Sensor Instrument Handbook, Ver. 5.0 (Baltimore, MD: STScI)
\bibitem[Kenyon et al. 1994]{ken94} Kenyon, S. J., Dobrzycka, D. \& Hartmann, L. 1994, AJ, 108, 1872
\bibitem[Lamzin et al. 2001]{lam01} Lamzin, S. A., Melnikov, S. Y., Grankin, K. N., \& Ezhkova, O. V. 2001, A\&A, 372, 922
\bibitem[Lattanzi et al. 1992]{lat01} Lattanzi, M. G., Bucciarelli, B., Holfeltz, S. T., \& Taff, L. G. 1992, in ASP Conf. Ser. 32, IAU Colloq. 135, Complementary Approaches to Double and Multiple Star Research, ed. H. A. McAlister \& W. I. Hartkopf (San Francisco: ASP), 377
\bibitem[Li et al. 2001]{li01} Li, J. Z., Ip, W. H., Chen, W. P., Hu, J. Y., \& Wei, J. Y.  2001, ApJ, 549, L89
\bibitem[Mason et al. 1999]{mas99} Mason, B. D., Douglass, G. G., \& Hartkopf, W. I. 1999, AJ, 117, 1023
\bibitem[Mason et al. 2001]{mas01} Mason, B. D., Wycoff, G. L., Hartkopf, W. I., Douglass, G. G., \& Worley, C. E. 2001, AJ, 122, 3466 (http://ad.usno.navy.mil/wds)
\bibitem[Nelan \& Makidon 2001]{fgsman} Nelan, E. P. \& Makidon, R. B. 2001, Fine Guidance Sensor Instrument Handbook for Cycle 11, Ver. 10 (Baltimore, MD: STScI)
\bibitem[Palla \& Stahler]{PS99} Palla, F. \& Stahler, S. W. 1999, ApJ, 525, 772
\bibitem[Simon et al. 1995]{S95} Simon, M. et al. 1995, ApJ, 443, 625
\bibitem[Simon et al. 1996]{S96} Simon, M., Holfeltz, S. T., \& Taff, L. G. 1996, ApJ, 469, 890
\bibitem[Simon et al. 2000]{S00} Simon, M., Dutrey, A., \& Guilloteau, S.  2000, ApJ, 545, 1034
\bibitem[Tamazian et al. 2002]{tam02} Tamazian, V. S., Docobo, J. A., White, R. J., \& Woitas, J. 2002, ApJ, 578, 925
\bibitem[Thi\'{e}baut et al. 1995]{T95} Thi\'{e}baut, E., Balega, Y., Balega, I., Belkine, I. Bouvier, J., Foy, R., Blazit, A., \& Bonneau, D. 1995, A\&A, 304, L17
\bibitem[White \& Ghez 2001]{WG01} White, R. J. \& Ghez, A. M. 2001, ApJ, 556, 265
\bibitem[Wizinowich et al. 2000]{aoref} Wizinowich, P. et al. 2000, Proc. SPIE, 4007, 64
\end{thebibliography}
\end{document}